\documentclass[twocolumn,aps,prd]{revtex4-1}
\usepackage{graphicx}
\usepackage{bm}
\usepackage{relsize}
\usepackage{amsmath,amsfonts,amsthm}
\begin{document}
\newcommand*{\bi}{\bibitem}
\newcommand*{\ea}{\textit{et al.}}
\newcommand*{\eg}{\textit{e.g.}}
\newcommand*{\zpc}[3]{Z.~Phys.~C \textbf{#1}, #2 (#3)}
\newcommand*{\plb}[3]{Phys.~Lett.~B \textbf{#1}, #2 (#3)}
\newcommand*{\phrc}[3]{Phys.~Rev.~C~\textbf{#1}, #2 (#3)}
\newcommand*{\phrd}[3]{Phys.~Rev.~D~\textbf{#1}, #2 (#3)}
\newcommand*{\phrl}[3]{Phys.~Rev.~Lett.~\textbf{#1}, #2 (#3)}
\newcommand*{\pr}[3]{Phys.~Rev.~\textbf{#1}, #2 (#3)}      
\newcommand*{\npa}[3]{Nucl.~Phys.~A \textbf{#1}, #2 (#3)}  
\newcommand*{\npb}[3]{Nucl.~Phys.~B \textbf{#1}, #2 (#3)}  
\newcommand*{\npbps}[3]{Nucl.~Phys.~B (Proc. Suppl.) \textbf{#1}, #2 (#3)}  
\newcommand*{\ptp}[3]{Prog. Theor. Phys. \textbf{#1}, #2 (#3)}
\newcommand*{\ppnp}[3]{Prog. Part. Nucl. Phys. \textbf{#1}, #2 (#3)}
\newcommand*{\ibid}[3]{\textit{ ibid.} \textbf{#1}, #2 (#3)}
\newcommand*{\epja}[3]{Eur. Phys. J. A \textbf{#1}, #2 (#3)}
\newcommand*{\epjc}[3]{Eur. Phys. J. C \textbf{#1}, #2 (#3)}
\newcommand*{\jpg}[3]{J. Phys. G \textbf{#1}, #2 (#3)}
\newcommand*{\mpla}[3]{Mod. Phys. Lett. A \textbf{#1}, #2 (#3)}
\newcommand*{\ijmpa}[3]{Int. J. Mod. Phys. A \textbf{#1}, #2 (#3)}
\newcommand*{\jhep}[3]{J. High Energy Phys. \textbf{#1}, #2 (#3)}
\newcommand*{\sjnp}[4]{Yad. Fiz. \textbf{#1}, #2 (#3) [Sov. J. Nucl. Phys.
\textbf{#1}, #4 (#3)]}
\newcommand*{\ptep}[3]{Prog. Theor. Exp. Phys. \textbf{#1}, #2 (#3)}
\newcommand*{\phrep}[3]{Phys. Rep. \textbf{#1}, #2 (#3)}
\newcommand*{\ra}{\rightarrow}
\newcommand*{\pippim}{\pi^+\pi^-}
\newcommand*{\kpkm}{K^+K^-}
\newcommand*{\rf}[1]{(\ref{#1})}
\newcommand*{\be}{\begin{equation}}
\newcommand*{\ee}{\end{equation}}
\newcommand*{\bea}{\begin{eqnarray}}
\newcommand*{\eea}{\end{eqnarray}}
\newcommand*{\nl}{\nonumber \\}
\newcommand*{\rmd}{\mathrm d}
\newcommand*{\lplm}{\ell^+\ell^-}
\newcommand*{\vtopill}{V^0\to\pi^0\,\ell^+\ell^-}
\newcommand*{\ktopiee}{K^\pm\to\pi^\pm e^+e^-}
\newcommand*{\ktopimm}{K^\pm\to\pi^\pm\mu^+\mu^-}
\newcommand*{\ktopill}{K^\pm\to\pi^\pm\ell^+\ell^-}
\newcommand*{\kppiee }{K^+\to\pi^+ e^+e^-}
\newcommand*{\kppimm} {K^+\to\pi^+\mu^+\mu^-}
\newcommand*{\die}{e^+e^-}
\newcommand*{\dimu}{\mu^+\mu^-}
\newcommand*{\jj}{\mathrm i}
\newcommand*{\ndf}{\mathrm{NDF}}
\newcommand*{\mev}{\mathrm{~MeV}}
\newcommand*{\cndf}{\chi^2/\mathrm{NDF}}
\newcommand*{\minuit}{\texttt{MINUIT}~}
\newcommand*{\w}{\sqrt s}
\newcommand*{\e}[1]{{\mathrm e}^{#1}}
\newcommand*{\ie}{\textrm{i.e.}}
\newcommand*{\dek}[1]{\times10^{#1}}
\newcommand*{\pio}{\mathrm{A}_{2\pi}}
\newcommand*{\piop}{\mathrm{A}^\prime_{2\pi}}
\newcommand*{\gp}{{1{\mathrm s}}}
\newcommand*{\ep}{{2{\mathrm p}}}
\newcommand*{\oee}{\omega\to\pi^0\,e^+e^-}
\newcommand*{\pee}{\phi\to\pi^0\,e^+e^-}
\newcommand*{\omm}{\omega\to\pi^0\,\mu^+\mu^-}
\newcommand*{\ffsq}{\left|F_{V^0\pi^0}(M^2)\right|^2}
\newcommand{\der}[2]{\frac{{\mathrm d}#1}{{\mathrm d}#2}}
\newcommand{\dd}{{\mathrm d}}
\def\babar{\mbox{\slshape B\kern-0.1em{\smaller A}\kern-0.1em
    B\kern-0.1em{\smaller A\kern-0.2em R}}}

\title{Possible manifestation of the 2p pionium in particle physics processes}
\author{Peter Lichard}
\affiliation{
Institute of Physics and Research Centre for Computational Physics
and Data Processing, Silesian University in Opava, 746 01 Opava, 
Czech Republic\\
and\\
Institute of Experimental and Applied Physics,
Czech Technical University in Prague, 128 00 Prague, Czech Republic
}
\begin{abstract}
We suggest a few particle physics processes in which excited 2p
pionium $\piop$ may be observed. They include the $e^+e^-\to\pi^+\pi^-$ 
annihilation, the $V^0\to\pi^0\ell^+\ell^-$ and $K^\pm\to\pi^\pm\ell^+\ell^-$ 
($\ell=e,\mu$) decays, and the photoproduction of two neutral pions
from nucleons. We analyze available experimental data and find that
they, in some cases, indicate the presence of 2p pionium, but do not 
provide definite proof.
\end{abstract}
\maketitle
\section{Introduction}
The first thoughts about an atom composed of a positive pion and a negative 
pion (pionium, or $\pio$ in the present-day notation) appeared almost 
sixty years ago. Uretsky and Palfrey \cite{uretsky} assumed its existence and 
analyzed the possibilities of detecting it in the photoproduction off hydrogen 
target. Up to this time, such a process has not been observed. They also 
hypothesized about the possibility of decay $K^+\to\pi^+\pio$, which has
recently been observed in the experiment we mention below.

Pionium was discovered in 1993 at the 70 GeV proton synchrotron at 
Serpukhov, Russia \cite{pionium}. The $\pio$ atoms were produced in a
Ta target and in the same target they broke-up into their constituents
with approximately equal energies and small relative momenta, which
distinguished them from the ``free'' $\pi^+\pi^-$ pairs.

Using a similar method, the properties of ground-state pionium were 
intensively studied in 
the Dimeson Relativistic Atomic Complex (DIRAC) experiment at the CERN Proton
Synchrotron \cite{shortlife}. A careful analysis showed that the mean pionium
lifetime is $\tau=3.15^{+0.28}_{-0.26}\dek{-15}$~s. It decays into two neutral 
pions \cite{uretsky,other} and, to a much lesser extent, into two photons 
\cite{gasser08}. 

The NA48/2 \cite{NA48/2} experiment at CERN Super Proton Synchrotron (SPS)
observed a cusplike structure in the $\pi^0\pi^0$ invariant mass
distribution from $K^\pm\to\pi^\pm\pi^0\pi^0$ decay. The enhancement can 
be interpreted as the contribution from the decays $K^\pm\to\pi^\pm\pio$ 
(considered in \cite{uretsky}) followed by the decay
$\pio\to\pi^0\pi^0$ \cite{cabibbo}.

The DIRAC collaboration recently discovered \cite{longlife} so-called 
long-lived $\pippim$ atoms. These objects are apparently excited 2p states of 
the ground-state pionium $\pio$. The discovery was enabled by modifying the
original DIRAC setup by adding a Pt foil downstream of the production Be
target. The breakup of the long-lived states happened in that foil, placed
at a distance of 96~mm behind the target. The magnetic field between the target
and the foil does not influence the path of neutral atoms, but the $\pippim$
pairs coming from various sources are made more divergent.

The longevity of 2p pionium ($\piop$) is caused by the fact that              
its quantum numbers $J^{PC}=1^{--}$ prevent it from decaying into the positive
C-parity $\pi^0\pi^0$ and $\gamma\gamma$ states. It must first undergo the
2p$\to$1s transition to the ground state. The mean lifetime of 2p
pionium 
\be
\label{longtau}  
\tau_{2\mathrm p}=0.45^{+1.08}_{-0.30}\dek{-11}~\mathrm s. 
\ee
is close to the value which comes for the $\pippim$ atom assuming a pure
Coulomb interaction \cite{longlife}. After reaching the 1s state, a decay 
to two $\pi^0$s quickly follows: 
$\pio^\prime\to\pio+\gamma\to\pi^0\pi^0\gamma$.

The quantum numbers of $A_{2\pi}^\prime$ allow its coupling to the 
electromagnetic field. Therefore, it can mediate, like the $\rho^0$
meson, the interaction of neutral hadronic systems with that field or with
a $C\!=\!-1$ system of charged leptons and photons. However, in contrast to 
the $\rho^0$ meson, the coupling to the photon of which is fixed by the
hypothesis of vector-meson dominance (VMD) \cite{vmd}, the coupling of
$\piop$ to the photon is unknown \cite{coupling}. In addition, the width of
$\piop$ is extremely narrow. It comes out as 1.46$\dek{-10}$~MeV if we 
take the central value of the measured lifetime \rf{longtau}.

In this paper, we will elaborate on some consequences of the fact that $\piop$
interacts with the electromagnetic field. To this end, we need an
estimate of the $\piop$ mass, which is related to the binding energy $b$ by
$M=2m_{\pi^+}-b$. Assuming pure Coulombic interaction, the binding energy 
of pionium can be calculated from the hydrogen-atom formula 
\be
\label{hatom}
b_n=\frac{m_r\alpha^2}{2n^2},
\ee
where $m_r$ is the reduced mass in energy units (used throughout this paper), 
$\alpha\approx 1/137.036$ is the fine-structure constant, and $n$ is the 
principal quantum number. Putting $n=2$ for the first excited state and 
$m_r=m_{\pi^+}/2$, we get $b=0.4645$~keV. The strong interactions may
shift the energies given by Eq.~\rf{hatom} \cite{gasser09}. As far as we know, 
there is no experimental or theoretical indication that the binding energies 
of the ground ($n=1$) or first excited state ($n=2$) differ significantly 
from their Coulombic values \rf{hatom}. Nevertheless, Uretsky and Palfrey
considered binding energies even higher than 10~MeV in their 
analysis~\cite{uretsky}. Today we know that pionium decays into two neutral 
pions, so the binding energy $b$ must be smaller than $2(m_{\pi^+}-m_{\pi^0})
\approx9.19$~MeV. 

In this paper we will consider, besides the Coulombic value $b=0.4645$~keV, 
the binding energy of $b=9$~MeV. We hope that the phenomena we are going 
to study will be able to decide between these two extreme values.

\section{Possible manifestation of  2$\bm p$ pionium in the
$\bm{\die\to\pippim}$ process}
Recently, we have succeeded \cite{kaonium} in locating 2p kaonium as a 
bound-state pole in the amplitude of the $\die\to K^+K^-$ process. 
The pole corresponding to 2p kaonium lies on the real axis in the complex 
$s$-plane below the $\kpkm$ threshold. Our aim here is to find a pole 
in $\die\to\pippim$ amplitude that would correspond to 2p pionium by 
fitting the data on the $\die\to\pippim$ cross section. Our experience with
2p kaonium shows that the crucial role in discovering the bound-state pole,
which lies below the reaction threshold, is played by the cross section
data at low energies, as close to the threshold as possible \cite{kaonium}.
Unfortunately, almost all $\die\to\pi^+\pi^-$ experiments have concentrated on 
the $\rho$/$\omega$ region or on energies above $\phi(1020)$. The only 
exception is the \babar~experiment \cite{babar2012}, which in 2012 covered 
a wide energy range from 0.3 to 3.0~GeV by exploring the initial-state 
radiation method \cite{isr}. The files containing the cross section data and 
their covariance matrices are provided in the Supplemental Material repository
(Ref. 32 in \cite{babar2012}).

To fit the \babar~cross-section data, we use the VMD formula for the cross 
section of the $\die$ annihilation into a pseudoscalar meson and its 
antiparticle with $n$ interfering resonances in the intermediate state  
\be
\label{sigma}
\sigma(s)=\frac{\pi\alpha^2}{3s}\left(1-\frac{4m_P^2}{s}\right)^{\!\frac{3}{2}}
\left|\sum_{i=1}^n
\frac{R_ie^{\jj\delta_i}}{s-M_i^2-\jj M_i\Gamma_i}
\right|^2.
\ee                                                            
Here, $M_i$ and $\Gamma_i$ determine the position and width of the $i$th
resonance, respectively. The residuum $R_i$ includes the product of two
constants. One characterizes the coupling of the $i$th resonance to the 
photon (up to the elementary charge $e$, which is taken off to form, 
after squaring, an $\alpha$ in the prefactor) and the other is the coupling 
of the resonance to the pseudoscalar meson pair. The phases $\delta_i$ 
regulate the interference between the resonances. We put $\delta_1=0$.

We first perform our fit over the full \babar~energy range assuming five 
``standard" resonances ($\rho$, $\omega$, $\rho^\prime$, 
$\rho^{\prime\prime}$, $\rho^{\prime\prime\prime}$) \cite{fred}.
Similarly as it was done in Ref. \cite{babar2012} when fitting the data on 
the pion form factor, we fit the cross-section data with only diagonal errors.
We get a perfect fit in terms of standard $\chi^2$ and the number of free
parameters (NDF): $\cndf=236.7$/318, which implies the confidence level
(C.L.) of 100\%. This is not good news from the point of searching for
pionium, because there is little room for improvement. The parameters
of the fit are shown in the second column of Table~\ref{tab:babar}.

Then we add two free parameters: the residuum of the assumed 2p pionium $R_6$ 
and the phase shift $\delta_6$. The width $\Gamma_6$ is set to zero because
the $\piop$ is a stable entity from the point of view of the
$\die\to\pi^+\pi^-$ reaction.
The 2p pionium mass is expressed by means of its binding energy $b$:
$M_2=2m_{\pi^+}-b$. We consider two cases: (i) $b=0.464$~keV (Coulombic
binding energy), (ii) $b=9$~MeV. The corresponding
fit parameters are shown in the third and fourth column of
Table~\ref{tab:babar}, respectively. In both cases, the drop of the $\chi^2$
against the case without pionium is more than 3, which with the two more
free parameters means that the fitting function has been changed in a sound
way. However, it cannot be considered proof of  2p pionium existence
because a perfect fit also exists without it.

\begin{figure}[h]
\includegraphics[width=8.6cm]{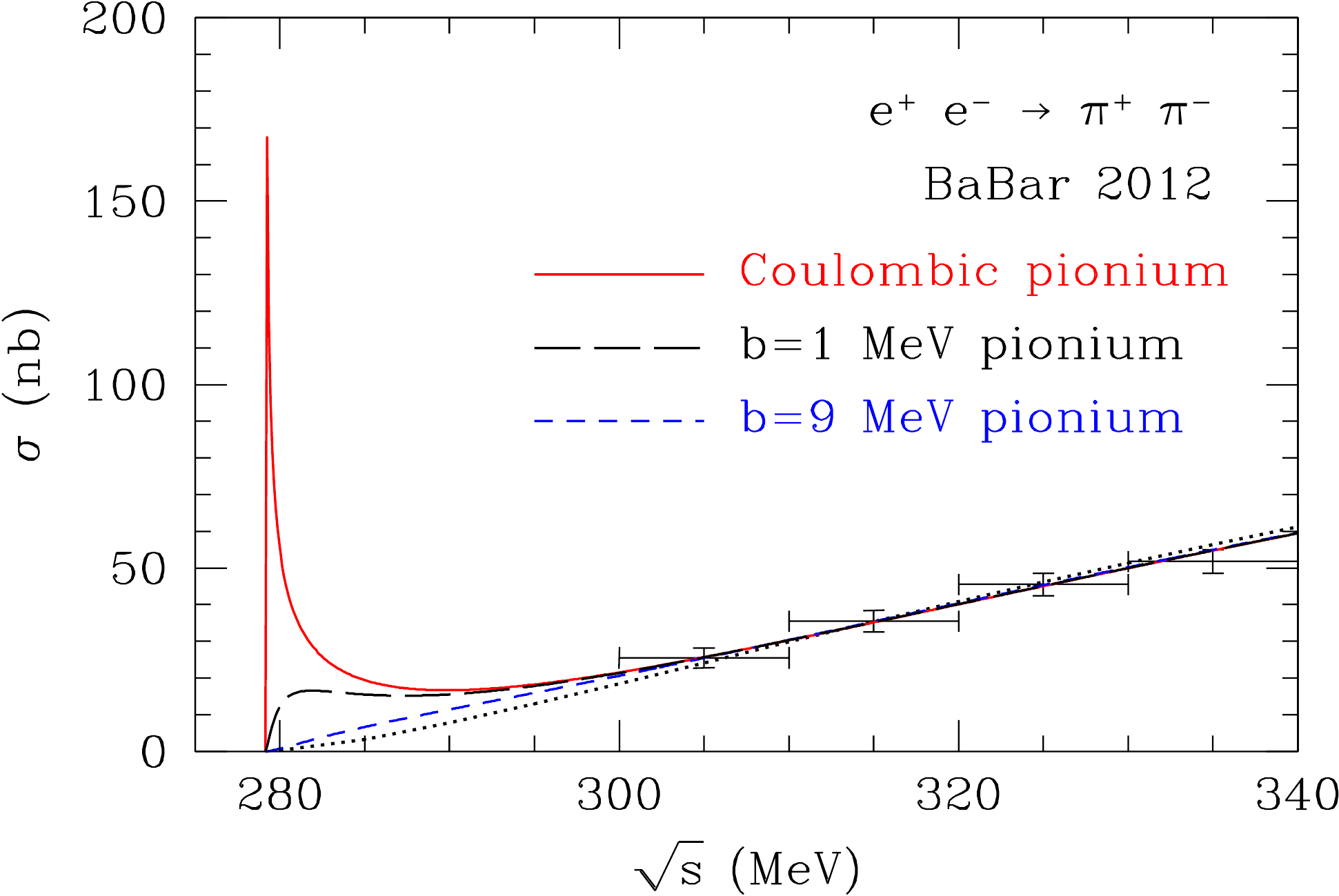}
\caption{\label{fig:fiveres30pionium}Low energy parts of the
\babar~\cite{babar2012} data and of our fit to all (0.3--3.0~GeV) 
data points. (a) dotted curve: the fit assuming five ``standard" resonances;
(b) full curve: resonances+Coulombic pionium; (c) long-dash curve:
resonances+pionium, $b\!=\!1$~MeV; (d) dashed curve: resonances+pionium,
$b\!=\!9$~MeV. Fit parameters are shown (except for the $b\!=\!1$~MeV case)
in Table \ref{tab:babar}.}
\end{figure}

However, even in this situation, something can be learned from the behavior
of the calculated cross sections at very low energies depicted in 
Fig.~\ref{fig:fiveres30pionium}. In the case of $b=9$~MeV, the excitation
curve differs very little from that without pionium and there would be
little chance to confirm pionium existence even if the data below 290~MeV were
available. A sure sign of the presence of Coulombic (or nearly
Coulombic) pionium would be if the lowest 10-MeV-wide bin were taller
than the next one. For intermediate binding energies, the sensitivity will
depend on the data precision. In Fig.~\ref{fig:fiveres30pionium},
a curve corresponding to $b=1$~MeV is also depicted.

The origin of the spike in Fig.~\ref{fig:fiveres30pionium} is obvious. The
pole corresponding to Coulombic pionium lies below the threshold, very close 
(0.464~keV) to it. So the beginning of the slope is already ``visible" 
above the threshold. In the amplitude squared, the rise continues to very
high values (to infinity as the $\Gamma_6$ is set to zero)
at $\sqrt s=M_6<2m_{\pi^+}$. In the excitation function, the rise is cut off 
by the $p^{*3}_\pi$ factor in Eq. \rf{sigma} and a vanishing cross section is 
reached at the reaction threshold.
\begin{table}[b]
\caption{\label{tab:babar}Parameters of the fits to 
 \babar~$\pippim$ data~\cite{babar2012} over the full energy range assuming 
no pole below the threshold (second column) and the 2p pionium  with the 
Coulombic binding energy $b$ (third column) or with $b=9$~ MeV 
(last column).}
\begin{tabular*}{8.6cm}[b]{lccc}
\hline
                &~No pionium~&~$b=0.464$~keV~&~$b=9$~MeV~\\
\hline
$R_1$ (GeV$^2$) & 0.7086(31) & 0.7114(28)  & 0.7060(30)  \\
$M_1$ (GeV)     & 0.75629(18)& 0.75645(21) & 0.75646(21) \\
$\Gamma_1$ (GeV)& 0.14361(33)& 0.14374(36) & 0.14377(36) \\
$R_2\!\times\!10^3$\,(GeV$^2$) & 7.65(28)& 7.73(29) & 7.74(28) \\
$M_2$ (GeV)     & 0.78203(18)& 0.78204(18) & 0.78204(18) \\
$\Gamma_2$ (MeV)& 8.16(34)   & 8.23(35)    & 8.24(34)    \\
$\delta_2$      & -2.02(36)  & -2.02(37)   & -2.02(36)   \\
$R_3$ (GeV$^2$) & 0.477(17)  & 0.483(38)   & 0.476(38)   \\
$M_3$ (GeV)     & 1.426(12)  & 1.414(15)   & 1.413(15)   \\
$\Gamma_3$ (GeV)& 0.465(22)  & 0.456(24)   & 0.455(23)    \\
$\delta_3$     & 2.63(13)    & 2.71(13)    & 2.74(13)\\
$R_4$ (GeV$^2$) & 2.63(13)   & 0.333(40)   & 0.328(39)    \\
$M_4$ (GeV)     &1.817(18)   & 1.810(19)   & 1.809(18) \\
$\Gamma_4$ (GeV)& 0.353(25)  & 0.333(30)   & 0.330(29)   \\
$\delta_4$      & -1.58(20)  & -1.31(25)   & -1.26(25)   \\
$R_5$ (GeV$^2$) & 0.045(24)  & 0.033(18)   & 0.032(17)  \\
$M_5$ (GeV)     & 2.239(28)  & 2.239(23)   & 2.238(22)  \\
$\Gamma_5$ (GeV)& 0.166(90)  & 0.126(85)   & 0.122(78)    \\
$\delta_5$      &-0.94(41)   & -0.71(48)   & -0.66(46)   \\
$R_6$ (GeV$^2$) &            & 0.0099(33)  & 0.0125(43)  \\
$M_6$ (GeV)     &            & $2m_{\pi^+}\!-b$   & $2m_{\pi^+}\!-b$ \\
$\Gamma_6$ (GeV)&            & 0           &   0        \\
$\delta_6$      &            &  -1.566(64) & 1.141(68)   \\
$\cndf$         & 236.7/318  & 233.5/316   &  233.4/316  \\
 Confidence level & 100\%    & 100\%       & 100\% \\
\hline
\end{tabular*}
\end{table}

\section{Possible manifestation of 2$\bm p$ pionium in the
$\bm{V^0\to\pi^0\ell^+\ell^-}$ decays}
In this section, we will try to spot the traces of 2p pionium in the most
recent data on the $ \oee$~\cite{A2y2017}, $\omm$~\cite{NA60y2016}, and 
$\pee$~\cite{KLOEy2016} decays.

\subsection{Phenomenology}
The formula that expresses the differential decay width of a neutral vector 
meson to a neutral pion and the dilepton with invariant mass $M$ in terms 
of the decay width of its real photon alternative \cite{landsberg1985} can be 
cast in the form
\bea
\label{dGdMsq1}
\der{\Gamma(V^0\to\pi^0\,\lplm)}{M^2}&=&\left(\frac{p_{\gamma_M}}
{p_\gamma}\right)^{\!3}\Gamma(V^0\to\pi^0\,\gamma)\nl
&\times&
{\cal T}(M^2)\ffsq\ , 
\eea
where $p_{\gamma_M}$ ($p_\gamma$) is the dilepton (photon) momentum in the 
$V^0$ rest frame, $F_{V^0\pi^0}$ is the transition form factor, and
\be
\label{tfun}
{\cal T}(M^2)=\frac{\alpha}{3\pi M^2}\left(1+\frac{2m^2}{M^2}\right)
\left(1-\frac{4m^2}{M^2}\right)^{\!1/2}.
\ee
The last function is known \cite{krollwada,baierkhoze,acta} as providing the 
connection between the production of the dilepton and a fictitious
massive photon $\gamma_M$. In our case,
\be
\label{dGdMsq}
\der{\Gamma(V^0\to\pi^0\,\lplm)}{M^2}=\Gamma(V^0\to\pi^0\,\gamma_M)\times
{\cal T}(M^2). 
\ee
We can thus write the relation
\be
\label{ffsq}
\ffsq=\frac{\Gamma(V^0\to\pi^0\,\gamma_M)}{\Gamma(V^0\to\pi^0\,\gamma)}
\times\left(\frac{p_\gamma}{p_{\gamma_M}}\right)^{\!3},
\ee
which will be explored later. 

\subsection{The model}
There are many interesting theoretical approaches to the $V^0\to\pi^0\lplm$
phenomenon \cite{theories}. As we want to concentrate on the role of
2p pionium, we will use the simplest possibility~--~the VMD model, which
allows the easy and transparent inclusion of 2p pionium into the
game. The simplest Lagrangian of the $V^0\!\rho^0\pi^0$
interaction \cite{meissner}
\be
\label{lagvrhopi}
{\cal L}(x)=g_{V\!\rho\pi}\ \epsilon_{\mu\nu\alpha\beta}\ \partial^\mu
V^\nu(x)\ \partial^\alpha\rho^\beta(x)\ \pi(x)
\ee
and the electromagnetic-current--vector-field identity \cite{vmd}
\[
J_\mu(x)=-\frac{e}{g_\rho}m_\rho^2\ \rho_\mu(x)
\]
are used to calculate both decay widths on the right-hand side of 
Eq.~\rf{ffsq}. In this way we obtain the form factor of the VMD model
\be
\label{vmdff}
F_{V^0\pi^0}(s)=\frac{m_\rho^2}{m_\rho^2-s+\jj m_\rho\Gamma_\rho(s)},
\ee
where $s=M^2$. The energy-dependent total width of the $\rho^0$ is given, 
at energies we will use, by the $\rho^0\to\pi^+\pi^-$ decay width
\be
\label{gammarho}
\Gamma_\rho(s)=\Gamma_\rho\frac{m_\rho^2}{s}\left(\frac{s-4m_{\pi^+}^2}
{m_\rho^2-4m_{\pi^+}^2}\right)^{3/2}.
\ee
The $\Gamma_\rho$ is the decay width of the $\rho^0$ at its nominal mass
$m_\rho$.

However, it is known, and recent experiments 
\cite{A2y2017,NA60y2016,KLOEy2016} have confirmed, that the VMD model 
provides 
a very poor description of the experimental data on the $V^0\to\pi^0\lplm$ 
decays. We will therefore modify the VMD model in a way that was suggested in
Ref.~\cite{ratio}, namely, by taking into account the possible energy 
dependence of the $V^0\to\rho^0\pi^0$ vertex by replacing the coupling 
constant in Eq.~\rf{lagvrhopi} by a strong form factor. Inspired by the 
flux-tube-breaking model of Kokoski and Isgur (KI) \cite{kokoski}, we assume
that the strong form factor behavior is given by formula 
\be
\label{ki}
G_{V\!\rho\pi}(p^*)=g_{V\!\rho\pi}\times
\exp\left\{-\frac{{p^*}^2}{12\beta^2}\right\},
\ee
where $p^*$ is the momentum of either of the particles coming out of the
$V^0\to\rho^0\pi^0$ vertex in the $V^0$ rest frame. KI
estimated the value of parameter $\beta$ at 0.4~GeV. We will consider
it a free parameter.

Applying  formula \rf{ki} to the decay widths on the right-hand side of 
Eq.~\rf{ffsq}, we find that the $\ffsq$ acquires, in comparison with the 
standard VMD, an extra factor of
\[
K(s)=\exp\left\{\frac{p_\gamma^2-p_{\gamma_M}^2}{6\beta^2}\right\}.
\]
Using $p_{\gamma_M}^2=\lambda(m^2_{V^0},m^2_{\pi^0},s)/(4m^2_{V^0})$,
where the ``triangle" function is defined by 
\be
\label{lambda}
\lambda(x,y,z)=x^2+y^2+z^2-2xy-2xy-2yz, 
\ee
we end up with
\[
K(s)=\exp\left\{\frac{s(2m^2_{V^0}+2m^2_{\pi^0}-s)}
{24m^2_{V^0}\beta^2}\right\}.
\]
Finally, adding the 2p pionium interfering with $\rho^0$, we arrive at
the form factor squared of our model
\bea
\label{myffsq}
|F_{V^0\pi^0}(s)|^2&=&\frac{K(s)}{(1+\epsilon)^2} 
\left|\frac{m_\rho^2}{m_\rho^2-s+\jj m_\rho\Gamma_\rho(s)}\right.\nl
&+&\left.
\epsilon\ \frac{m_{2\pi}^2}{m_{2\pi}^2-s+\jj m_{2\pi}\Gamma_{2\pi}}
\right|^2.
\eea
The constant $\epsilon$ will be treated as a free parameter. 

The experimental data are provided as mean values of the form factor squared
within bins in $M$. We therefore produce the model outcome in the
same format. It must be said that due to the presence of an extremely narrow
resonance in the form factor \rf{myffsq} it would be unthinkable to proceed
in a different way, \eg, by calculating the form factor squared in 
isolated points. Another possible way would be to convolute the model outcome 
with the $M$-resolution of a particular experiment.

As the experimental bins are immensely wider than the 2p pionium width, 
the last contributes only to a single bin. For calculating the model outcome
in that bin, we use the numerical quadrature method described in the
Appendix.

\subsection{Results}
\subsubsection{$\oee$ decay}
The A2 Collaboration at Mainz Microtron (MAMI) in 2017 presented the first
measurement of the dielectron mass spectrum in the $\oee$ decay.
The electron beam from the MAMI-C accelerator produced Bremsstrahlung
photons in a radiator. The photons hit a liquid hydrogen target and
initiated the $\gamma p\to\omega p$ reaction. The results were
presented in the form of the $\omega\pi^0$ transition form factor squared 
$|F_{\omega\pi^0}|^2$ \cite{A2y2017}. Previously, the data on this quantity
were only obtained from the dimuon mass spectrum in the $\omm$ decay
\cite{NA60y2016,leptong,NA60y2009}. 

\begin{figure}[h]
\includegraphics[width=8.6cm]{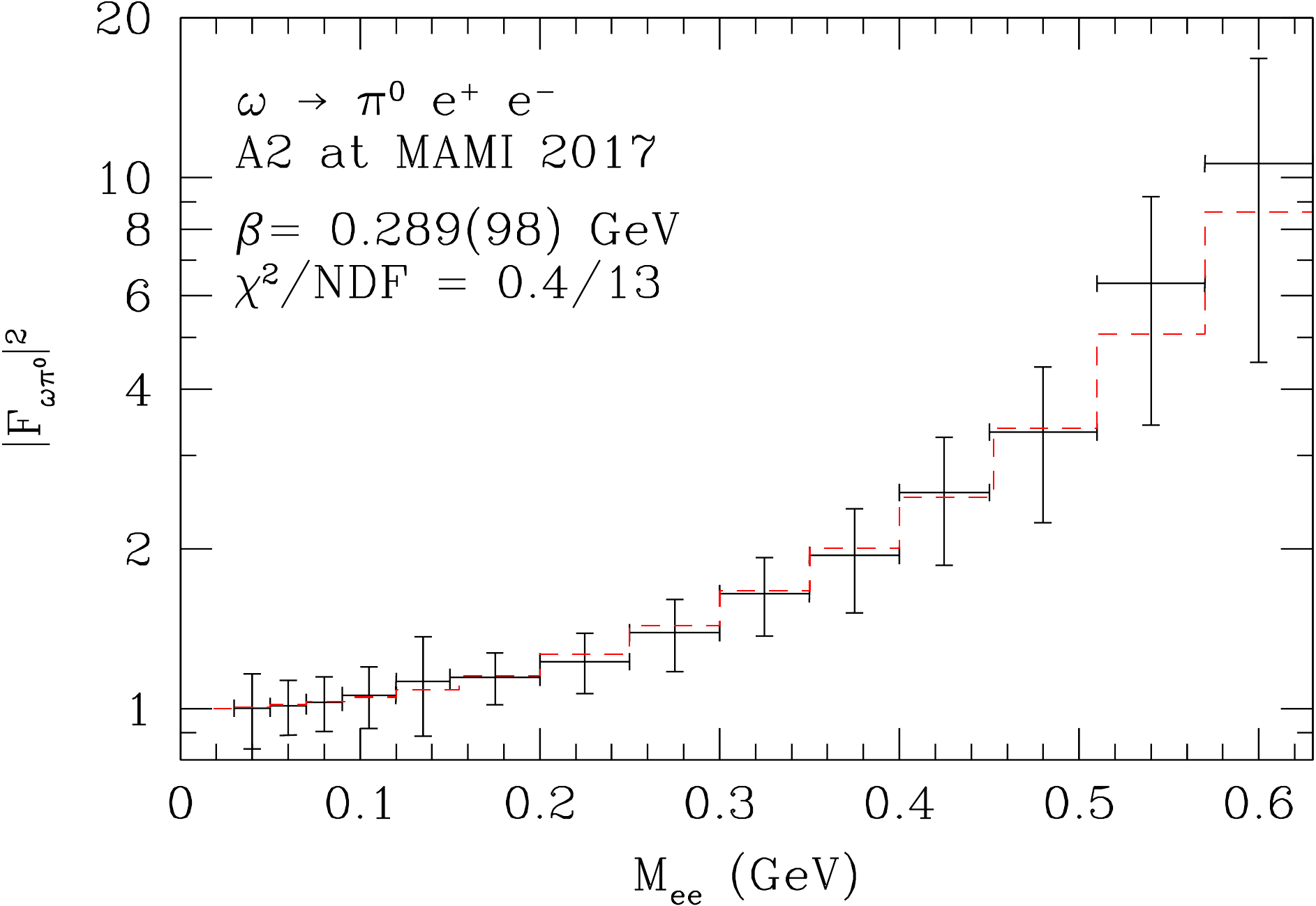}
\caption{\label{fig:A2ffsq}Fit of our model to the A2 Collaboration at MAMI
data \cite{A2y2017} on $\oee$ decay. The dashed histogram represents the 
outcome of the
model without the $\piop$ contribution ($\epsilon=0$) . As the model is 
above the data in the bin $[0.25,0.30]$~GeV, in which a possible contribution
from $\piop$ is expected, a nonvanishing $\epsilon$ cannot improve the fit
(because of the extreme narrowness and huge peak value of $\piop$, the 
destructive interference is excluded).}
\end{figure}

Using our model Ansatz \rf{myffsq} without the $\piop$ contribution
($\epsilon\equiv0$) we get a very good fit ($\cndf=0.4/13$) to the A2 data,
see Fig. \ref{fig:A2ffsq}. The optimal value of the parameter
$\beta$ is $0.289\pm0.098$~GeV, which is not far from the value of 0.4~GeV
suggested by Kokoski and Isgur \cite{kokoski}.
When considering our main interest in this paper, the manifestation of 2p
pionium, the A2 data are not supportive at all. As the mass of 2p pionium
must lie between $2m_\pi^0$ and $2m_\pi^+$, it would appear in the bin
[0.25~GeV,~0.30~GeV]. However, the model value of $|F_{\omega\pi^0}|^2$ in this
bin is already higher than the experimental one. Therefore, including 2p 
pionium can only worsen the fit quality. 

\subsubsection{$\pee$}
The KLOE experiment at the DA$\Phi$NE $\die$ collider at Frascati, Italy,
is the site where historically the first measurement of the dielectron 
mass distribution in the $\pee$ decay was performed. The results in the form 
of the $\phi\pi^0$ transition form factor were presented by KLOE-2 
Collaboration in Ref. \cite{KLOEy2016}.

\begin{figure}[h]
\includegraphics[width=8.6cm]{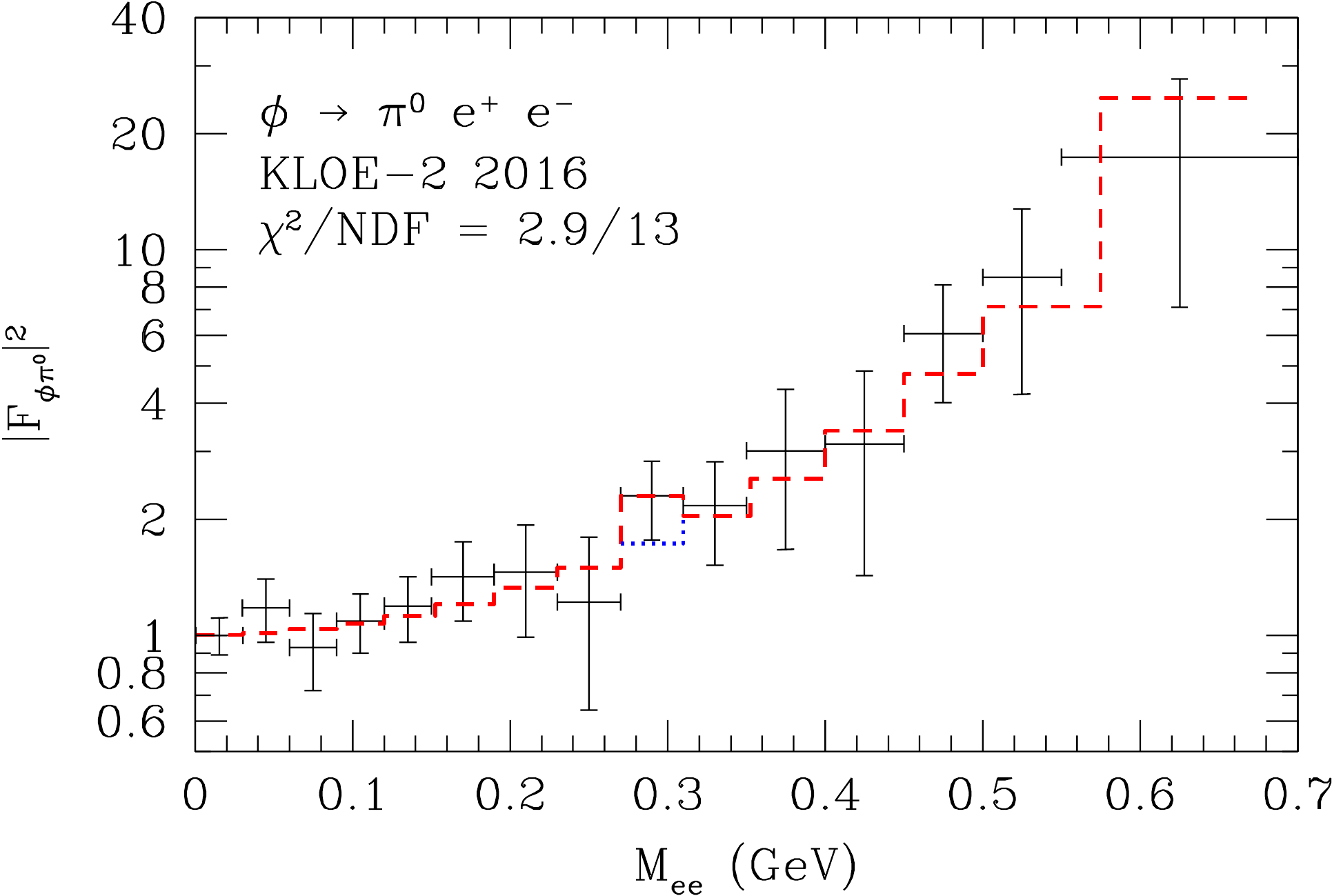}
\caption{\label{fig:KLOEffsq_full}Comparison of our model with the KLOE-2
data \cite{KLOEy2016}. Model outcome (dashed histogram) is the same for both 
assumed values of the 2p pionium binding energy because both alternative 
$\piop$ masses fall in the same $[0.27,0.31]$~GeV bin. In this bin, the
model perfectly matches the experimental value thanks to the fitting parameter
$\epsilon$, which regulates the contribution of $\piop$ to the form factor
squared. The dots in the $[0.27,0.31]$~GeV bin indicate the 
form-factor-squared value without the 2p pionium contribution.}
\end{figure}

We fit the KLOE-2 data by our form factor formula \rf{myffsq} using three 
options: (i) no 2p pionium ($\epsilon=0$), (ii) 2p pionium with the
Coulombic binding energy of 0.464~keV, and (iii) 2p pionium with the binding
energy of 9~MeV. The results are listed in Table \ref{tab:kloe2016} and 
displayed in Fig. \ref{fig:KLOEffsq_full}.
\begin{table}[h]
\caption{\label{tab:kloe2016}Parameters of the fits of our model to the
KLOE-2 Collaboration data \cite{KLOEy2016} on $\pee$ decay.}
\begin{tabular*}{8.6cm}[b]{lccc}
\hline
                &No pionium&~~$b=0.464$~keV~~&~~$b=9$~MeV~~\\
\hline
$\epsilon\dek{7}$ &0 (fixed)  & 1.49(72)  & 1.55(74)  \\    
$\beta$~(GeV)     & 0.168(21) & 0.174(25) & 0.174(25) \\
$\cndf$           & 4.0/14    & 2.9/13    & 2.9/13    \\
 Confidence level & 99.5\%    & 99.8\%    & 99.8\%    \\
\hline
\end{tabular*}
\end{table}
The calculated dependence of $|F_{\phi\pi^0}|^2$ on dielectron mass is the
same for all three options with one exception: in the bin extending from
0.27 to 0.31~GeV the two options with 2p pionium match the experimental value,
whereas the option without $\piop$ is somewhat lower. This may be considered
an indication of the role of 2p pionium in the $\pee$ decay.

\begin{figure}[h]
\includegraphics[width=8.6cm]{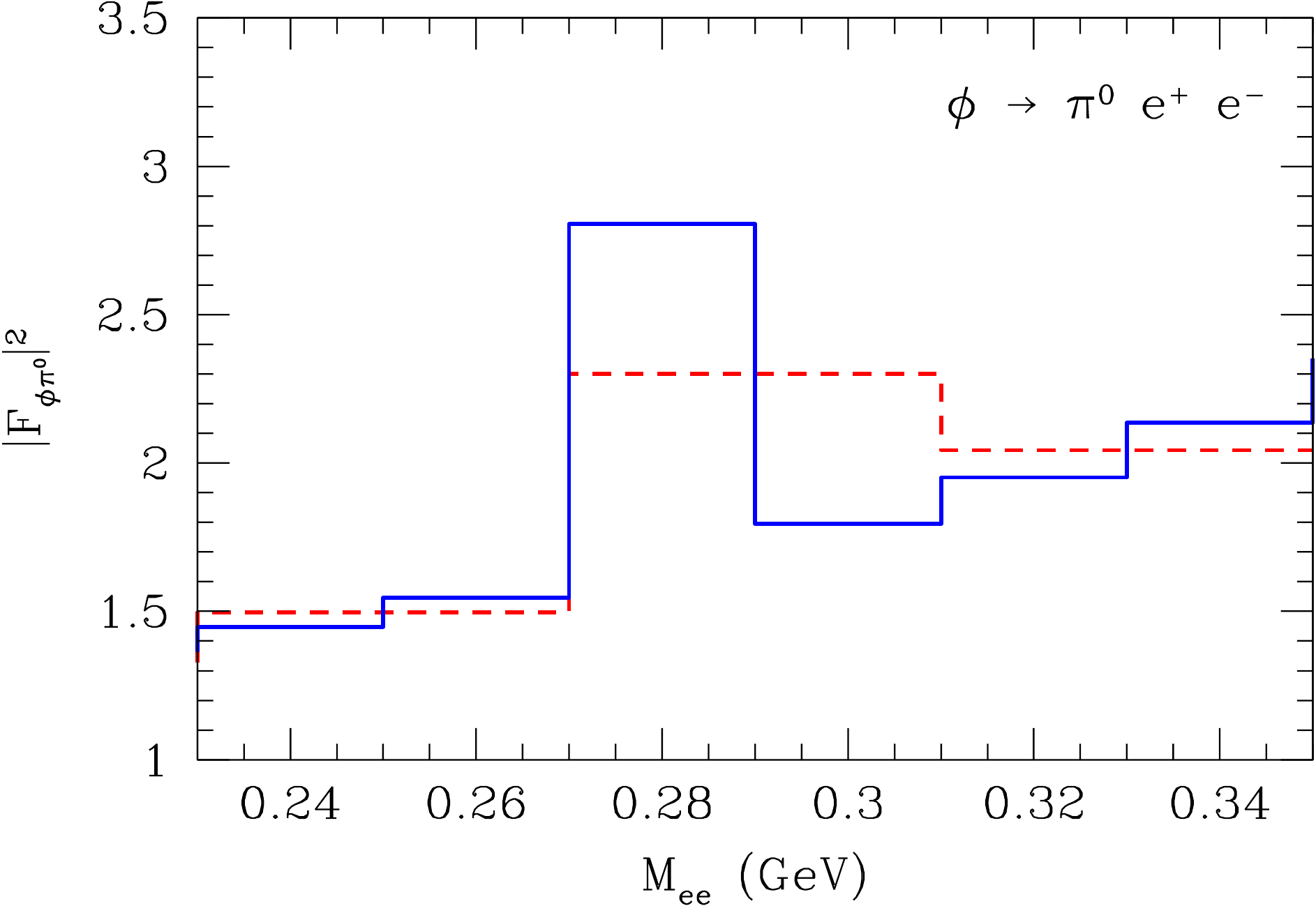}
\caption{\label{fig:KLOEffsq_detail}The original histogram (a part of that
shown in Fig. \ref{fig:KLOEffsq_full}) (dashed) and a histogram with
narrower bins (full).
 Illustrating the idea that by dividing
a bin into two sub-bins one can decide whether it contains the contribution 
from a narrow resonance or not. Of the three displayed bins, splitting 
the middle one provides two sub-bins with very different contents because
the resonance position falls into the left sub-bin. In two side bins, the
difference between sub-bins is not so pronounced, they do not contain a
resonance.}
\end{figure}

However, the enhancement of a single bin in comparison with its neighbors
may be a statistical fluctuation. The way to decide whether the
enhancement is a real effect or just a fluctuation is to use narrower bins, 
if the statistics permit. The idea is illustrated in
Fig.~\ref{fig:KLOEffsq_detail}. The imbalance between the two sub-bins is 
more significant if the original bin contains a $\piop$. In our
example, 2p pionium falls into the left sub-bin $[0.27,0.29]$~GeV for both 
choices of the binding energy. In a high-statistics experiment, it would
be possible to get the pionium mass better localized by choosing sufficiently
narrow bins. 

\subsubsection{$\omm$}
The NA60 Collaboration studied the $\omm$ decays in two experiments
performed at the CERN SPS. In the first experiment
\cite{NA60y2009}, the $\omega$ mesons were produced in 158$A$ GeV In-In
collisions. In the second one \cite{NA60y2016}, a system of nine subtargets 
of different nuclear species was exposed to an incident 400 GeV proton beam.  
The results of the two experiments are compatible \cite{NA60y2016}. We will
only use the results of the latter experiment, presented as the mean values
of the $\omega\pi^0$ transition form factor squared in 20~MeV-wide dimuon
mass bins.

Unfortunately, our model is not able to fit the data over the whole range of
dimuon masses
from 0.20 to 0.64 GeV. Not only our model, but also all the models investigated
in Ref. \cite{NA60y2016} are unable to follow a steep rise of the form
factor above 0.5 GeV found in both NA60 experiments. A similar rise
was reported by the Lepton G experiment \cite{leptong} performed at the
Institute for High Energy Physics, Serpukhov, Russia, in 1981.

To have a good fit in the region around the possible occurrence of the
$\piop$, we do not include the dimuon masses greater than 0.48~GeV in
our fits. The parameters of our three fits (no pionium, Coulombic binding,
binding energy of 9 MeV) are shown in Table~\ref{tab:na60}. 
\begin{table}[h]
\caption{\label{tab:na60}Parameters of the fits of our model to the
NA60 Collaboration data \cite{NA60y2016} on $\omm$ decay for $M_{\mu\mu}<
0.48$~GeV.}
\begin{tabular*}{8.6cm}[b]{lccc}
\hline
                &No pionium&~~$b=0.464$~keV~~&~~$b=9$~MeV~~\\
\hline
$\epsilon\dek{7}$ &0 (fixed) & 0.63(52) & 0.65(54) \\    
$\beta$~(GeV)     & 0.227(24)& 0.228(25)& 0.228(25)\\
$\cndf$           & 7.5/13   & 7.1/12   & 7.1/12  \\
 Confidence level & 87.5\%   & 85.1\%   & 85.1\% \\
\hline
\end{tabular*}
\end{table}

\begin{figure}[h]
\includegraphics[width=8.6cm]{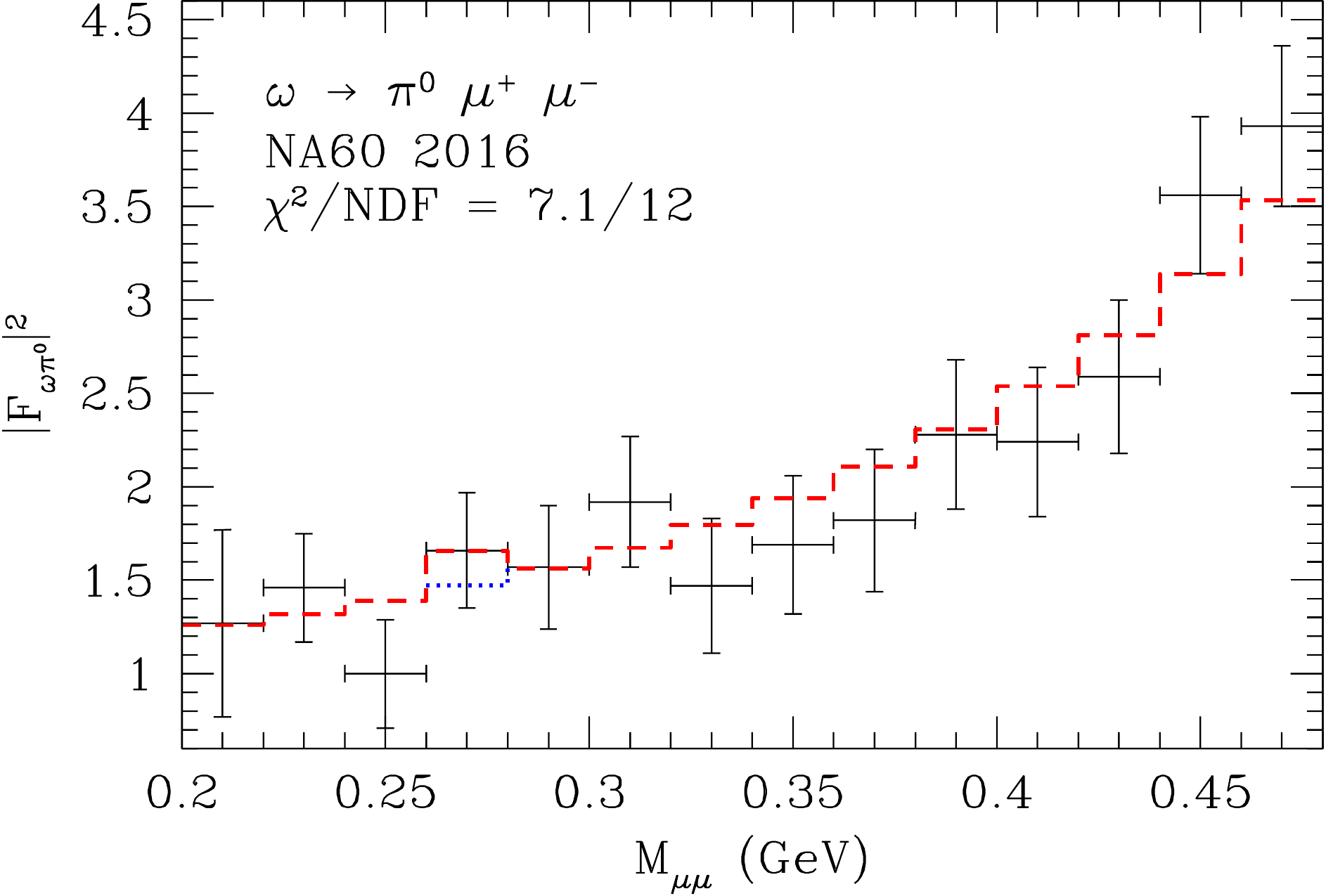}
\caption{\label{fig:NA60ffsq_14bins}Comparison of our model with the NA60
\cite{NA60y2016} data. The model outcome (dashed histogram) is the same for 
both assumed values of the 2p pionium binding energy because both alternative 
$\piop$ masses fall in the same $[0.26,0.28]$~GeV bin. In this bin, the
model perfectly matches the experimental value thanks to the fitting parameter
$\epsilon$, which regulates the contribution of $\piop$ to the form factor
squared. The dots in the $[0.26,0.28]$~GeV bin indicate the form factor
squared value without the 2p pionium contribution.}
\end{figure}

When producing a graphical output (Fig. \ref{fig:NA60ffsq_14bins}) our task
has again been facilitated by the fact that outside the $[0.26,0.28]$~GeV
bin all three fits give the same histogram. In that bin, the two 2p pionium 
options give identical results, the option without pionium gives a somewhat 
lower value.

\begin{figure}[h]
\includegraphics[width=8.6cm]{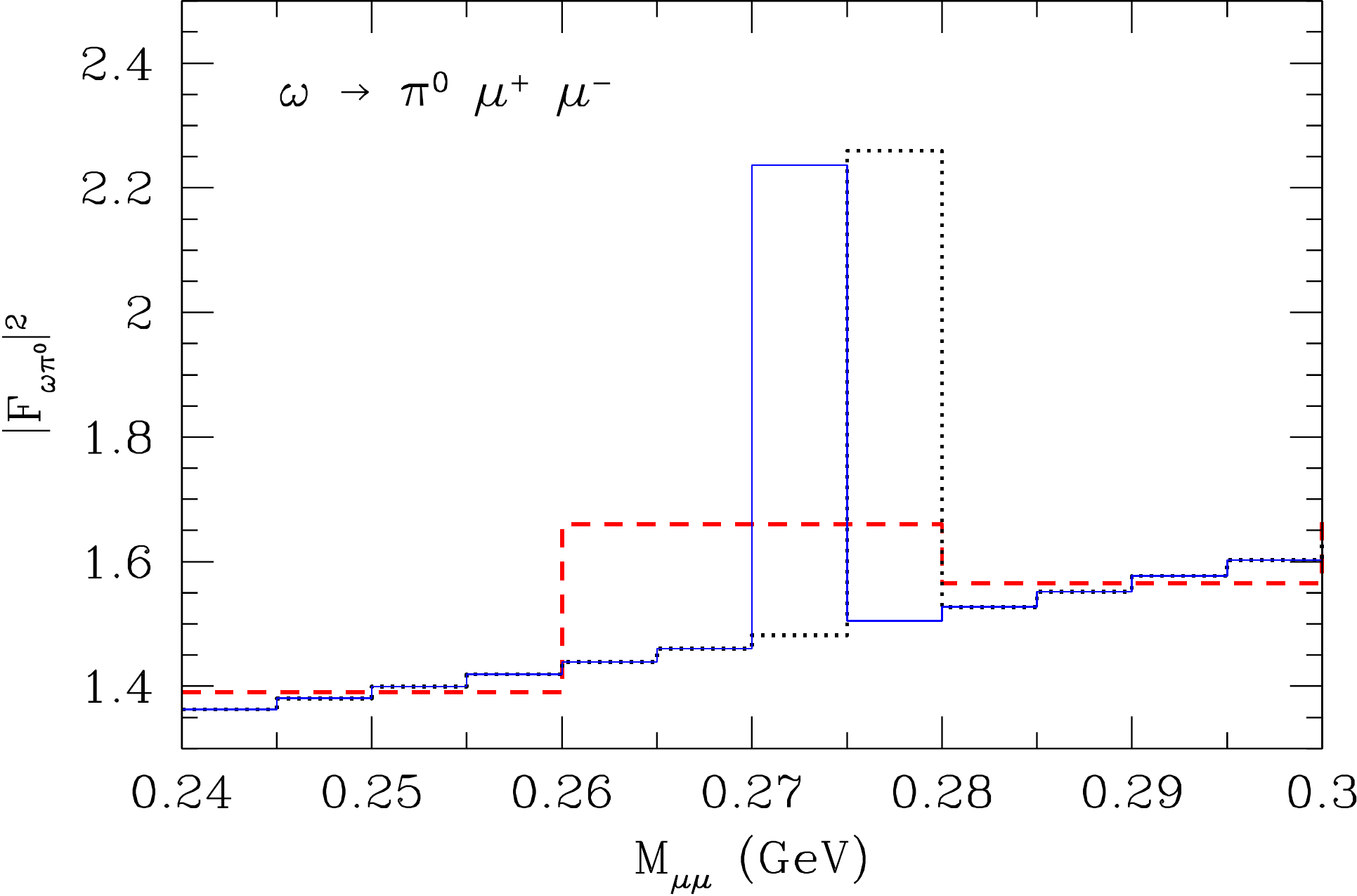}
\caption{\label{fig:NA60ffsq_ddetail}Dashed line: A part of the original
histogram shown in Fig. \ref{fig:NA60ffsq_14bins}); Solid line: the
calculation assuming a 9~MeV binding energy of 2p pionium; Dotted line:
2p pionium with Coulombic binding energy. Parameters $\beta$ and $\epsilon$
are kept at values shown in Table~\ref{tab:na60}.}
\end{figure}

The bin width in the NA60 data is 20~MeV. To model what would happen if a
narrower bin width were chosen, we first set it to 10~MeV and got a similar
picture as in the KLOE-2 case (Fig.~\ref{fig:KLOEffsq_detail}), just the role
of sub-bins was exchanged (in the NA60 case the right sub-bin is taller than
the left one). More interesting is the case with the bin width of 5~MeV,
see Fig.~\ref{fig:NA60ffsq_ddetail}. Now, two binding energy options produce
two different histograms. The mass of the Coulomb-bound $\piop$ is in
the $[0.275,0.280]$~GeV bin, whereas the mass of that with $b=9$~MeV
lies in the $[0.270,0.275]$~GeV bin.

\section{Possible manifestation of 2$\bm p$ pionium in the
$\bm{K^\pm\to\pi^\pm\ell^+\ell^-}$ decays}
The decays $K^\pm\to\pi^\pm\ell^+\ell^-$ have been examined in several
experiments \cite{kpill}. We will use the $\die$ and $\dimu$ data from the 
NA48/2 experiment \cite{batley09,batley11}, which are available in tabular 
form.

Below, we will show that the NA48/2 experiment does not provide any indication 
about the existence of the $\piop$ with Coulombic binding. The evidence of
the existence of $\piop$ with a higher binding energy (we consider 9~MeV, but 
somewhat smaller values are also possible) is stronger.

\subsection{Phenomenology}
The experimental data on the differential decay rate are most often
presented in terms of the dimensionless variable $z=M^2/m_K^2$, where
$M$ is the invariant dilepton mass, $M^2=(p_{\ell^+}+p_{\ell^-})^2$, and
$m_K$ is the charged kaon mass \cite{evelina}.
To get the model yield in a particular $z$-bin, one needs to calculate
the integral of $\dd \Gamma/\dd z$. In the presence of a narrow resonance,
it is more convenient to calculate the integral of $\dd \Gamma/\dd M$
(see Appendix) and use
\[
\int_{z_1}^{z_2}\der{\Gamma}{z}\dd z=\int_{M(z_1)}^{M(z_2)}\der{\Gamma}{M}\dd M
\]
Using the model-independent formula for $\dd\Gamma/\dd z$ valid in the 
one-photon approximation \cite{ecker,ambrosio} and the relation 
$\dd{z}/\dd{M}=2M/m_K^2$, we arrive at the formula
\bea
\label{dGammadM}
\der{\Gamma}{M}&=&\frac{G_F^2\alpha^2m_K^3}{6\pi(4\pi)^4}M\lambda^{\frac 3 2}
(1,M^2/m_K^2,m_\pi^2/m_K^2)\nl
&\times&\sqrt{1-4\frac{m_\ell^2}{M^2}}\left(1+2\frac{m_\ell^2}{M^2}\right)
\left|f(M^2)\right|^2,
\eea
where function $\lambda$ is defined in Eq. \rf{lambda}. A  particular
model is defined by specifying the (unnormalized) form factor $f(M^2)$.
The prefactor on the right-hand side of Eq. \rf{dGammadM} is chosen in such 
a way that $f(0)=f_0$, where $f_0$ is the parameter used  by 
the NA48/2 Collaboration in their papers \cite{batley09,batley11}.

\subsection{Model}
\begin{figure}[h]
\includegraphics[width=8.6cm]{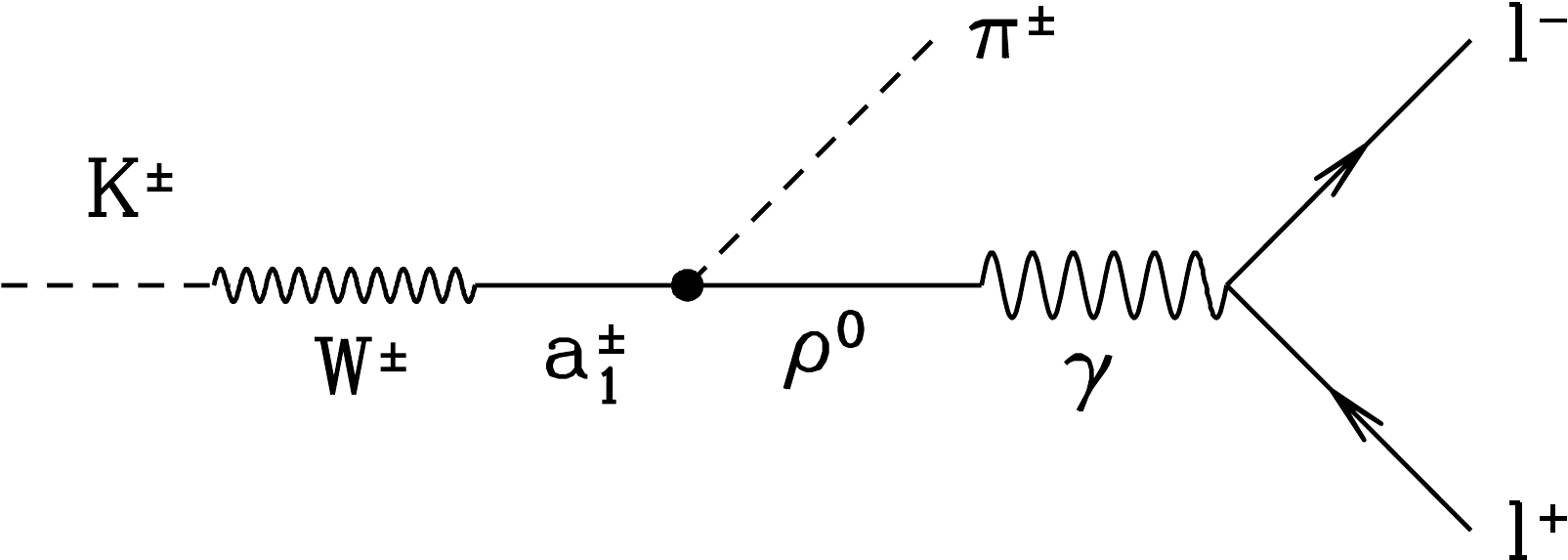}
\caption{\label{fig:kpill}Meson dominance model \cite{md} of the 
$\ktopill$ decay.}
\end{figure}

We will use a model based on meson dominance (MD) hypothesis \cite{md} 
depicted in Fig.~\ref{fig:kpill}, supplemented with  2p pionium 
interfering with the $\rho^0$ in the intermediate state. 
It was shown that the MD model can provide a reasonable estimate of the
$\kppiee$ decay rate using the information about the
$\tau^+\to\bar\nu_\tau\pi^+\pi^+\pi^-$ and $K^+\to\mu^+\nu_\mu$ decay rates
\cite{md,ratio}. But it failed in explaining the $\kppimm$ to $\kppiee$
branching ratio and the dilepton mass distribution shape, even when the
KI strong form factor \cite{kokoski} was taken into account 
\cite{ratio}. Also here, we will replace the $a_1\rho\pi$ coupling constant 
by an energy-dependent form factor
\be
\label{kikpill}
G_{a_1\!\rho\pi}(p^*)=g_{a_1\!\rho\pi}\times
\exp\left\{-\frac{{p^*}^2}{12\beta^2}\right\},
\ee
where $p^*$ is the 3-momentum of either of the particles coming out of the
$a_1\to\rho\pi$ vertex in the $a_1$ rest frame. However, there is a catch.
The relation \rf{kikpill} from the flux-tube breaking model \cite{kokoski}
is valid when the parent meson is on the mass shell $p_{a_1}^2=m_{a_1}^2$.
In the MD model of the $\ktopill$ decay, visualized in Fig.~\ref{fig:kpill}, 
the $a_1$ meson is off the mass shell, $p_{a_1}^2=m_K^2$. We will make an
additional assumption that the relation \rf{kikpill} holds also in this 
case. It results in diminishing the form factor in \rf{dGammadM} by ($s=M^2$)
\be
\label{lfaktor}
L(s)=\exp\left\{\frac{s(2m^2_K+2m^2_\pi-s)}
{48m^2_K\beta^2}\right\}.
\ee
Contrary to Ref.~\cite{ratio}, we will not use the KI value of 
$\beta=0.4$~GeV (valid for the on-mass-shell parent mesons) but will consider 
$\beta$ a free parameter. Adding the $\piop$ interfering with the $\rho^0$ 
in the intermediate state and putting all pieces together we get
\be
f(s)=f_0F(s),
\ee
with the normalized form factor equal to
\bea
\label{myff}
F(s)&=&\frac{L(s)}{(1+\epsilon)} 
\left[\frac{m_\rho^2}{m_\rho^2-s+\jj m_\rho\Gamma_\rho(s)}\right.\nl
&+&\left.
\epsilon\ \frac{m_{2\pi}^2}{m_{2\pi}^2-s+\jj m_{2\pi}\Gamma_{2\pi}}
\right].
\eea

\subsection{Results}

\subsubsection{$\ktopiee$}
The NA48/2 experiment at the CERN SPS used simultaneous
$K^+$ and $K^-$ beams produced by 400~GeV$/c$ proton impinging on a
beryllium target. After momentum selection and focusing, the beams entered
the fiducial decay volume with a length of 114~m. The decay products were
registered and measured by the very complex NA48 detector. The results based 
on the data set collected in 2003-2004 included the rates, spectra, and 
charge asymmetry. They were presented in 2009 \cite{batley09}.
We will use the dielectron $z$-spectrum in tabular form, which can be found
in \cite{durpdgna48ee}.    

\begin{figure}[h]
\includegraphics[width=8.6cm]{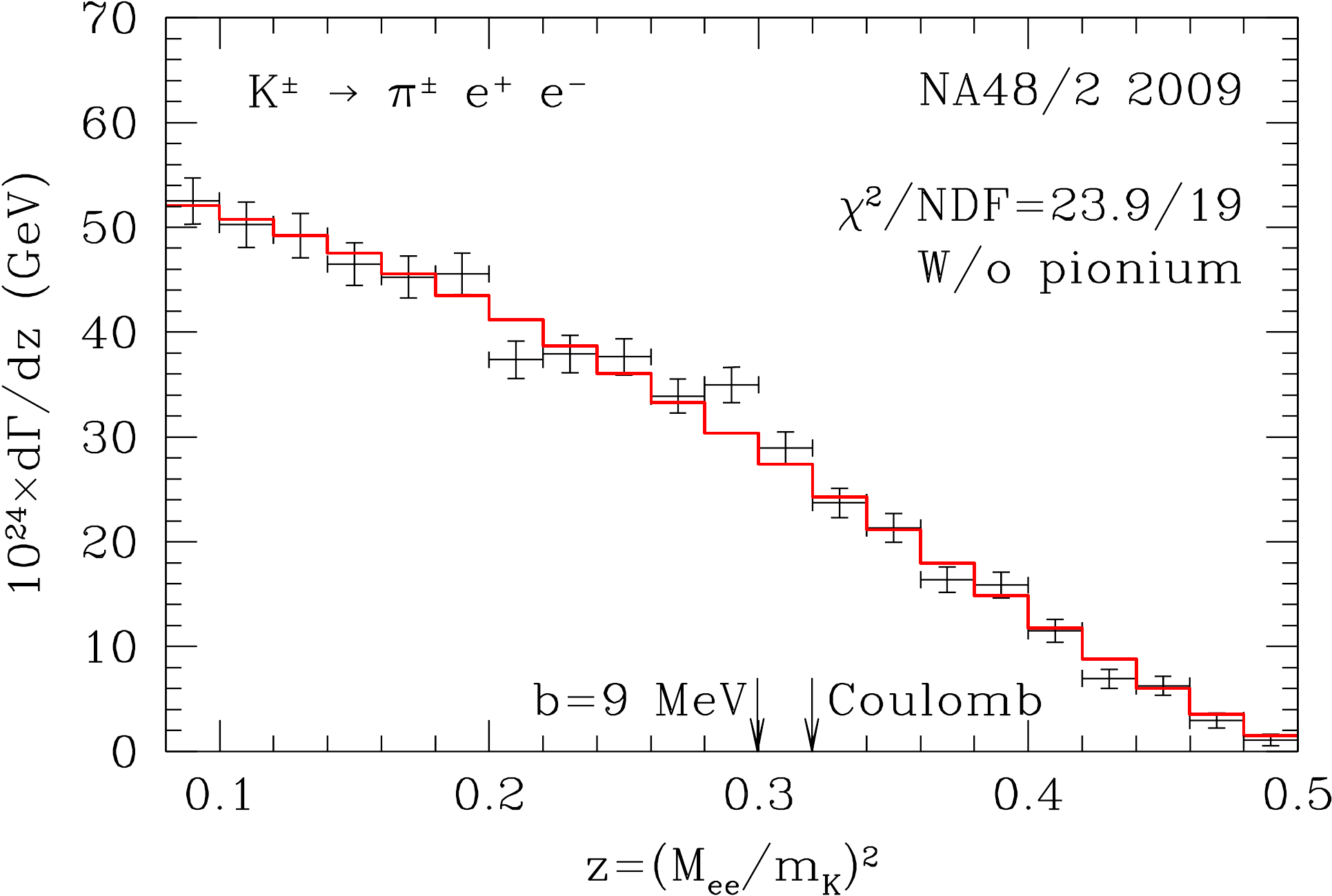}
\caption{\label{fig:eezdistr}Fit of the modified MD model to the NA48/2
data. No pionium is included. The arrows show the assumed $z$-positions of 
the 2p pionium with binding energy of 9~MeV and of that with Coulombic binding 
energy.}
\end{figure}

\begin{figure}[h]
\includegraphics[width=8.6cm]{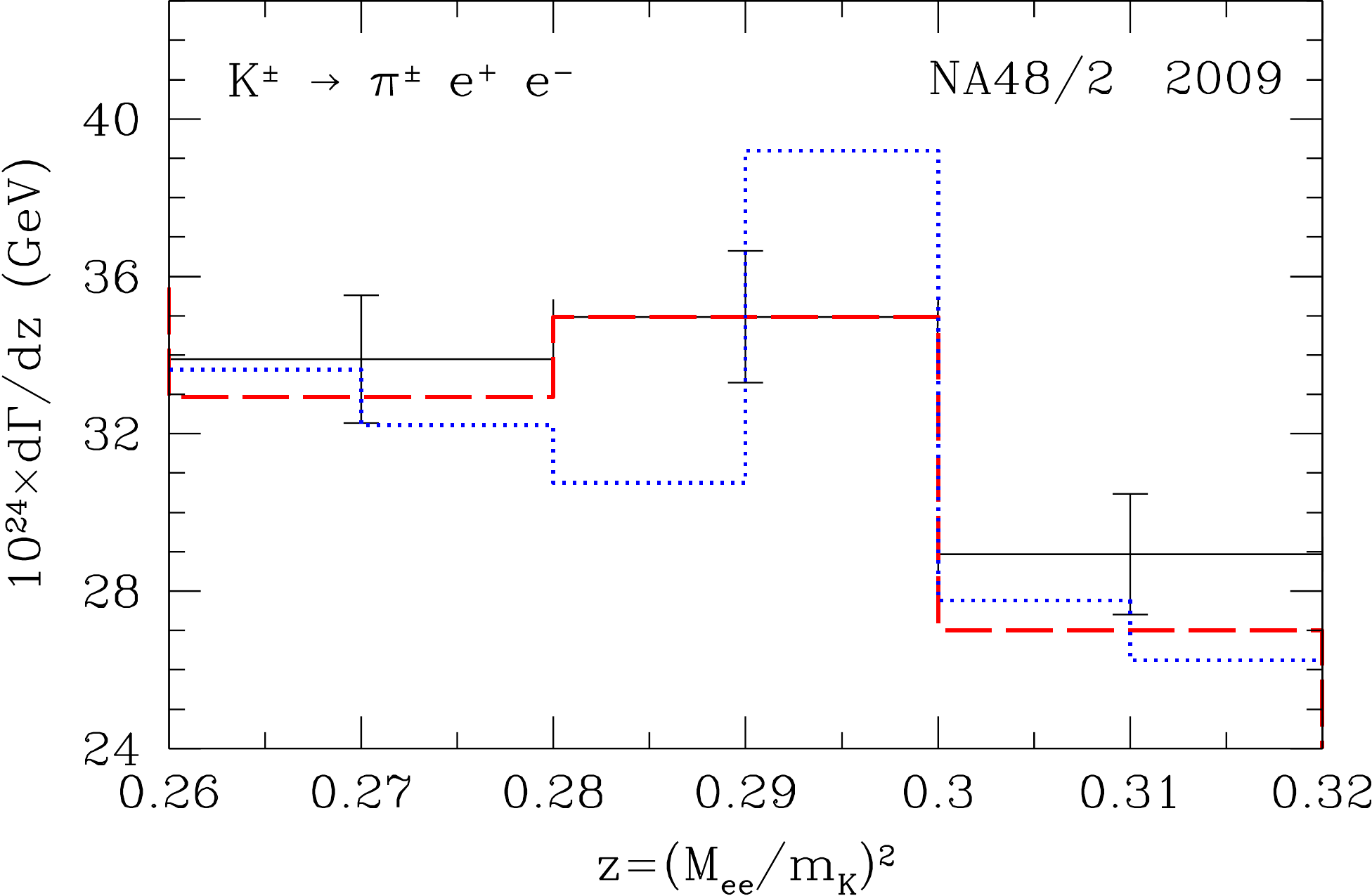}
\caption{\label{fig:eezdistr_ku1}Detail of the fit to the NA48/2 data
after the 2p pionium with binding energy of 9~MeV has been included
(dashed histogram). The model expectation with narrower bins is also
shown (dotted histogram).}
\end{figure}

We first fit the full $z$-spectrum (21 points) with the MD model without 
2p pionium ($\epsilon\equiv0$). See the leftmost data column in Table 
\ref{tab:eezdistr} and Fig. \ref{fig:eezdistr}. 
\begin{table}[h]
\caption{\label{tab:eezdistr}Results of the fits of the MD model 
to the NA48/2 data \cite{batley09} on $\ktopiee$ decay.}
\begin{tabular*}{8.6cm}[b]{lccc}
\hline
                  &~No pionium~&~$b=0.464$~keV~&~$b=9$~MeV~\\
\hline
$\epsilon\dek{7}$ & 0 (fixed)  & 0.30(14)    & 0.512(91) \\    
$\beta$~(GeV)     & 0.0909(28) & 0.0916(30)  & 0.0924(30)\\  
$ f_0$            & 0.5622(96) & 0.5635(97)  & 0.5646(98)\\        
$\cndf$           & 23.9/19    & 22.7/18     & 15.8/18  \\
 Confidence level & 20.0\%     & 20.2\%      & 60.6\% \\
\hline
\end{tabular*}
\end{table}
After including any of the two 2p pioniums, the histogram remains the same
as in Fig. \ref{fig:eezdistr} except the bin where a particular pionium 
sits. In that bin, a perfect match with data is reached by varying parameter
$\epsilon$. As the confidence levels in Table \ref{tab:eezdistr} show, the 
NA48/2 data prefer the 2p pionium with higher binding energy. This is also
visible in Fig. \ref{fig:eezdistr}, where the data point is well above
the model in bin $z\in[0.28, 0.30]$. In Fig. \ref{fig:eezdistr_ku1} we 
show the detailed histogram stemming from the model with the $b=9$~MeV 2p 
pionium added. The histogram with the same bin width 
as in experimental data is shown by a dashed line (now, the experimental value
is exactly matched). The dotted line shows the calculated half-width histogram.
It suggests the way experimentalists may decide whether an excess over 
the model without pioniums (or in comparison with neighboring bins) is a 
statistical fluctuation or a sign of 2p pionium presence. 

The contribution of a narrow resonance does not depend much on its exact $z$
position within a bin. Inspecting Fig. \ref{fig:eezdistr_ku1} we can say
that any 2p pionium with a $z$ position less than 0.3 would have the same
effect as that with $b=9$~MeV. In terms of binding energy it means that
any $\piop$ with $b>8.78$~MeV are acceptable. When taking into account
the higher bound given by the $m_{\pi^+}-m_{\pi^0}$ difference, we can
say that the $\die$ data of the NA48/2 Collaboration confine the 2p 
pionium binding energy to a narrow interval $b\in(8.78,9.19)$~MeV.

\subsubsection{$\ktopimm$}
The measurement of the $\ktopimm$ decay based on the data collected by the
NA48/2 experiment at the CERN SPS was reported in Ref. \cite{batley11}.
The numerical values of the presented $z$-distribution are available at
\cite{durpdgna48mm}.

\begin{figure}[h]
\includegraphics[width=8.6cm]{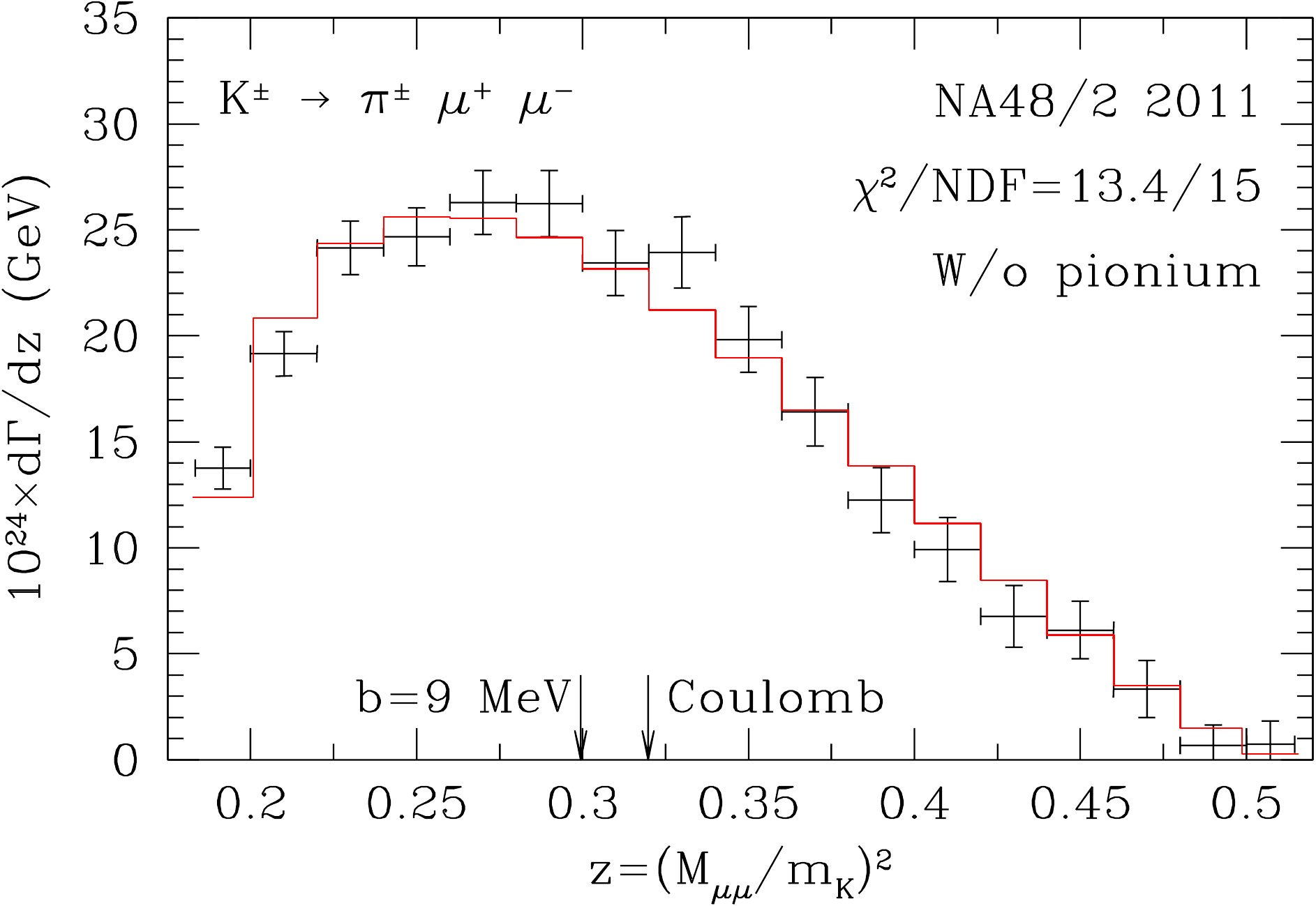}
\caption{\label{fig:mmzdistr}Fit of the modified MD model to the NA48/2
data. No pionium is included. The arrows show the assumed $z$-positions of
the 2p pionium with binding energy of 9~MeV and that
with Coulombic binding energy.}
\end{figure}

The results of our fits are 
shown in Table \ref{tab:mmzdistr}. The model without any pionium
($\epsilon=0$) exhibits a very good fit (C.L. of 62.5\%). The histogram 
showing the corresponding $z$-distribution is depicted in 
Fig.~\ref{fig:mmzdistr}. 

In the bin $z\in[0.30,0.32]$, where the Coulombic 
$\piop$ resides, the model and experimental values almost coincide. There is 
no room for the improvement of $\chi^2$ by allowing nonvanishing $\epsilon$.
Thus, the $\dimu$ mode does not allow the presence of the 2p pionium with
Coulombic binding energy. 
\begin{table}[h]
\caption{\label{tab:mmzdistr}Results of the fits of the modified MD model to 
the NA48/2 data \cite{batley11} on $\ktopimm$ decay.}
\begin{tabular*}{8.6cm}[b]{lccc}
\hline
                &~No pionium~&~$b=0.464$~keV~&~$b=9$~MeV~\\
\hline
$\epsilon\dek{7}$ & 0 (fixed)  & 0.14(36)   & 0.33(16)  \\
$\beta$~(GeV)     & 0.0857(61) & 0.0859(63) & 0.0863(63)\\  
$f_0$             & 0.544(28)  & 0.544(29)  & 0.545(29) \\        
$\cndf$           & 13.4/15    & 13.4/14    & 12.3/14   \\
 Confidence level & 57.1\%     & 49.5\%     & 58.2\%    \\
\hline
\end{tabular*}
\end{table}

\begin{figure}[h]
\includegraphics[width=8.6cm]{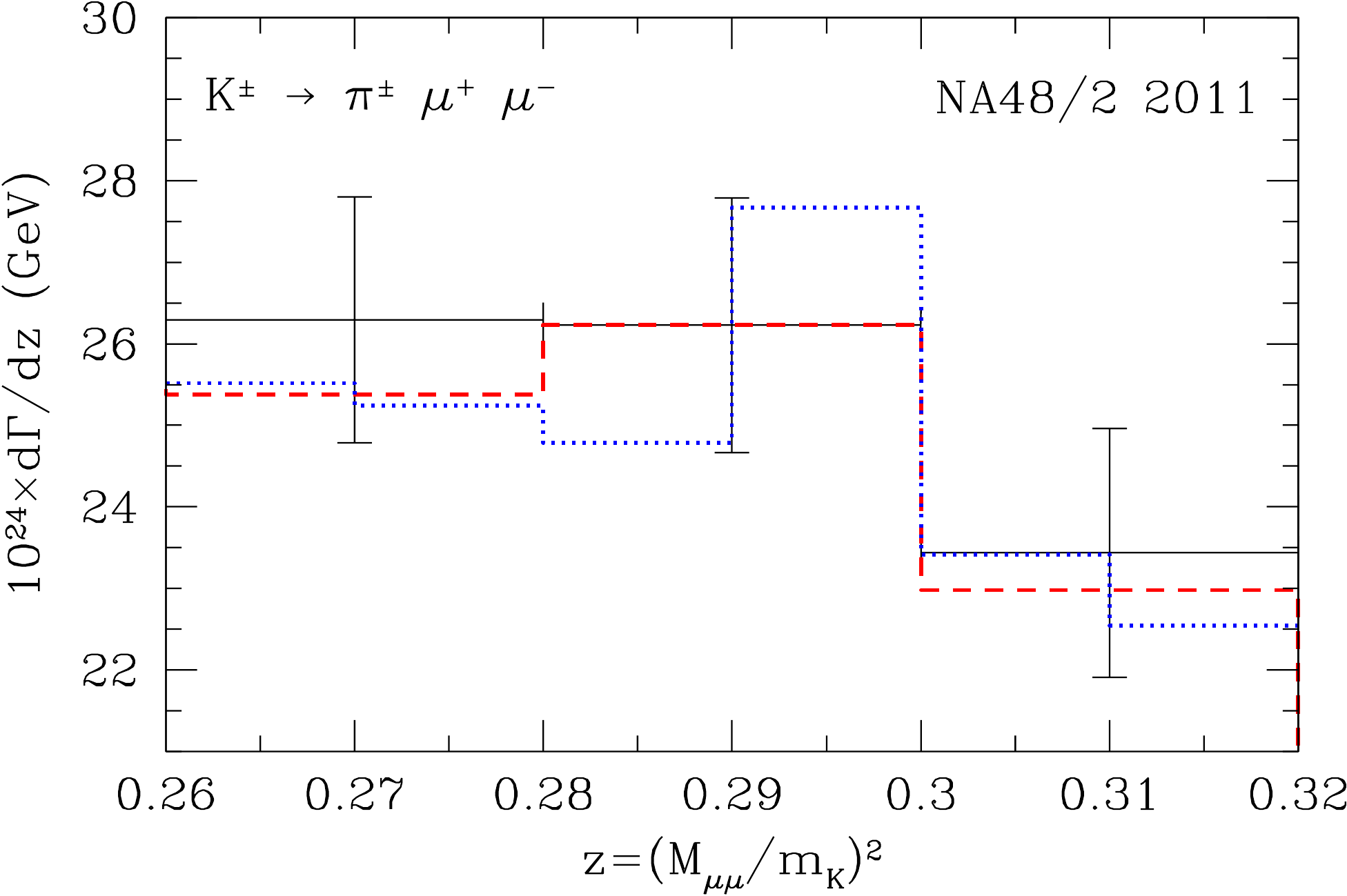}
\caption{\label{fig:mmzdistr_ku1}Detail of the fit to the NA48/2 data
after the 2p pionium with binding energy of 9~MeV has been included
(dashed histogram). The model expectation with narrower bins is also
shown (dotted histogram).}
\end{figure}

The $\piop$ with $b=9$~MeV is less salient than in the $\die$ mode.
Its inclusion improves the confidence level only marginally. Additionally,
the value of the parameter that characterizes the admixture of $\piop$ in the 
form factor \rf{myff} ($\epsilon=0.33\pm0.16$) is smaller than that in 
Table~\ref{tab:eezdistr} ($0.512\pm0.091$). Nevertheless, 
Fig.~\ref{fig:mmzdistr_ku1} again illustrates that narrower bins may help 
to identify the $z$-position of the resonance.

\section{Photoproduction of $\bm\piop$ from nucleons}
A long time ago \cite{uretsky}, the photoproduction reaction
\[
\gamma +p\to b^0 +p
\]
was considered a convenient means of detecting the scalar $\pippim$ atom
$b^0$ ($\pio$, in today's notation). None of the numerous experiments,
starting with the bubble chamber experiments at Cambridge, DESY, and at SLAC
in the 1960s, and continuing with many electronic experiments to the
present day (see  \cite{photo2020} for a list of the most recent ones), has 
had a glimpse of pionium. Most probably this is because  ground-state pionium 
does not couple to an electromagnetic field. On the other hand, 2p pionium 
can be produced
in inelastic Compton scattering, where it directly couples to the outgoing
virtual photon. However, 2p pionium exhibits special properties, which
should be taken into account. Its most salient feature is the long lifetime
\rf{longtau}, which means a long decay length. For example, if an event
is initiated by a photon with energy $E_\gamma=1$~GeV, the maximum 
$\piop$ momentum is 0.959~GeV/$c$ and the corresponding decay length is 
4.6~mm. So, the $\piop$ events will have a two-vertex structure, with 
separation between vertices of a few millimeters. In the first vertex,
the target proton gets a kick against the emitted 2p pionium, while two 
$\pi^0$s and a very soft photon appear from the decay chain 
$\piop\to\pio+\gamma\to\pi^0\pi^0\gamma$ in the 
second vertex. The four-momenta of the four energetic $\gamma$ quanta 
(coming from two $\pi^0$'s) should combine to the invariant
mass of $\pio$, which is a little below $2m_{\pi^+}$.

Even if the current photoproduction experiments are devoted to the study
of nucleon resonances, the by-product of an independent confirmation of 
the existence of long-lived pionium, discovered in \cite{longlife}, and 
getting its lifetime would be very valuable.

\section{Conclusions}
The message from available experimental data about the appearance of
2p pionium is mixed: 

The $\die\to\pippim$ process, even if supplemented with the cross section 
data closer to the threshold, would only be able to reject
or confirm a low-binding-energy ($\lesssim 1$~MeV) $\piop$, but would not 
be able to say anything about those with higher binding energy.
 
As of $V^0\to\pi^0\ell^+\ell^-$ decays, the A2 Collaboration data
\cite{A2y2017} on $\oee$ decay provide no room for $\piop$. The
experimental value of the form factor squared in the bin where 2p pionium
is expected is higher than the model result without pionium. 
What concerns the decays $\omm$ (NA60 \cite{NA60y2016} and $\pee$(KLOE-2 
\cite{KLOEy2016}) they exhibit very similar behavior. For each of them, 
the confidence levels of the three fit options (no pionium, Coulombic pionium,
pionium with $b=9$~MeV) are the same. There is some room for pionium with
any binding energy, the parameter $\epsilon$, which measures the $\piop$
contribution to the form factor is not vanishing, but with low statistical
significance. 

The strongest indication of the $\piop$ presence is provided by the
$\ktopiee$ data \cite{batley09} of NA48/2 Collaboration. While the fit with 
no pionium gives C.L. of 20.2\%, the inclusion of 2p pionium with 
$b=9$~MeV increases C.L. to 60.6\% and leads to
$\epsilon=(0.512\pm0.091)\dek{-7}$. Evidence for the Coulombic bound $\piop$
is very weak. Similar, but not so impressive, results are obtained by
analyzing the NA48/2 data on the $\ktopimm$ process \cite{batley11}.

The photoproduction data \cite{photo2020} exhibit a steep rise of the
$\pi^0\pi^0$ mass spectrum above the threshold, but it is unclear to which
extent it signals the presence of $\piop$. A dedicated analysis of the
measured data would be useful, taking into account the specific features
of possible $\piop$ production events. 

A final remark concerns two rare kaon decay experiments KOTO
\cite{koto} and NA62 \cite{na62}. Decays $K\to\pi\pio$ and $K\to\pi\piop$
may constitute a part of the background to the main investigated decay
$K\to\pi\nu\bar\nu$ \cite{kpinunu}. While the decay $K^\pm\to\pi^\pm\pio$ 
has already been discovered \cite{NA48/2}, the decays $K^\pm\to\pi^\pm\piop$, 
$K_L\to\pi^0\pio$, and $K_L\to\pi^0\piop$ are still awaiting  observation.
 
\begin{acknowledgments}
I am indebted to Evelina Mihova Gersabeck and Michal Koval for useful 
correspondence.

This work was partly supported by the Ministry of Education, Youth and
Sports of the Czech Republic Inter-Excellence Projects No. LTI17018
and No. LTT17018.
\end{acknowledgments}
\appendix*
\section{Integrating over the interval containing a narrow resonance}
\label{appendix}
When comparing the model results with the experimental data presented
as mean values over the finite bins we need to calculate the expressions
of the type
\be
I=\int_{W_1}^{W_2}\ g(W)\dd W
\ee
If the function $g(W)$ contains a narrow-resonance term
\[
\frac{1}{(W^2-M^2)^2+(M\Gamma)^2}
\]
($W_1<M<W_2$), the numerical quadrature may be rather erratic. Using the 
substitution
\[
W(\xi)=M+\frac{\Gamma}{2}\tan\left[\frac{a_2-a_1}{2}\xi+
\frac{a_2+a_1}{2}\right],
\]
where
\[
a_i=\arctan\frac{2(W_i-M)}{\Gamma}
\]
($i=1,2$) we arrive at the expression
\be
I=\frac{a_2-a_1}{\Gamma}\int_{-1}^1\
g[W(\xi)]\left[(W(\xi)-M)^2+\frac{\Gamma^2}{4}\right]\dd\xi,
\ee
which can be conveniently evaluated by the Gauss-Legendre method. It is
also convenient to split a wider bin to a part (preferably symmetric around
$M$) that contains the resonance and to the rest, which can be handled by 
standard quadrature methods.

\end{document}